\pdfoutput=1 
\documentclass[runningheads]{llncs}
\usepackage[T1]{fontenc}
\usepackage[utf8]{inputenc}
\DeclareUnicodeCharacter{2248}{\approx}
\DeclareUnicodeCharacter{2264}{\leq}
\DeclareUnicodeCharacter{2009}{\,}
\usepackage{subcaption}
\usepackage{booktabs}
\usepackage{graphicx}
\usepackage{amsmath}
\usepackage{enumitem}
\usepackage{amssymb}
\usepackage{algorithm}
\usepackage{algpseudocode}

\newif\ifpreprint
\preprinttrue        

\ifpreprint
  \usepackage{geometry}
\fi

\begin{document}
%
%

	\title{BlockSDN-VC: A SDN-Based Virtual Coordinate-Enhanced Transaction Broadcast Framework for High-Performance Blockchains}
	
%
\author{Wenyang Jia\inst{1} \and
Jingjing Wang\inst{1} \and
Ziwei Yan\inst{1} \and
Guohui Yuan\inst{1} \and
Tanren Liu\inst{2} \and
Yakun Ren\inst{2} \and
Kai Lei\orcidID{0000-0001-9197-895X}\inst{1}\thanks{Corresponding author.}}

\authorrunning{Jia, Lei et al.}
%
\institute{ICNLab, Shenzhen Graduate School, Peking University, Shenzhen, P.R.China\\ 
\and
SF Technology, Shenzhen, P.R.China\\
}
\maketitle              
\begin{abstract}
Modern blockchains need fast, reliable propagation to balance security and throughput. Virtual-coordinate methods speed dissemination but rely on slow iterative updates, leaving nodes out of sync. We present BlockSDN-VC, a transaction-broadcast protocol that centralises coordinate computation and forwarding control in an SDN controller, delivering global consistency, minimal path stretch and rapid response to churn or congestion. In geo-distributed simulations, BlockSDN-VC cuts median latency by up to 62 \% and accelerates convergence fourfold over state-of-the-art schemes with under 3 \% control-plane overhead. In a real blockchain environment, BlockSDN-VC boosts confirmed-transaction throughput by 17 \% under adversarial workloads, requiring no modifications to existing clients.

\end{abstract}

\section{INTRODUCTION}

Public blockchains must move each transaction from its point of origin to a quorum of geographically scattered validators before consensus can even begin; the slower this broadcast, the lower a system’s safe block-generation rate and the longer users must wait for confirmation. Classic overlays such as those in Bitcoin and Ethereum therefore sacrifice throughput for safety, achieving only dozens of transactions per second because their gossip layers need multiple round-trip times to converge. Recently, MERCURY showed that engineering the dissemination plane pays off: by embedding peers in a virtual coordinate system (VCS) and issuing an early-outburst fan-out, MERCURY cuts propagation delay by roughly half while adding little overhead.

We take a different approach. Because many high-throughput or consortium chains already operate in data-centre or ISP environments that run Software-Defined Networking (SDN), we externalise coordinate maintenance to the SDN control plane. The controller has a global, real-time map of paths and delays; it can compute consistent coordinates for every peer in one shot, push a sub-kilobyte relay table to each node, and send incremental updates whenever network conditions shift. This design eliminates iterative convergence delay, prevents coordinate forgeries, and lets the system reroute around congestion within a single control loop.

Building on this idea we propose BlockSDN-VC, a broadcast protocol that pairs controller-driven coordinates with a relay policy inspired by MERCURY’s early-outburst. A sender fans out to its d closest neighbours plus a small random cross-cluster sample; subsequent hops follow the same table, yielding logarithmic path length and diversity even if the controller becomes unreachable.

Packet-level simulations on a 1 000-node, geo-distributed topology show that BlockSDN-VC lowers the median propagation latency by up to 62 \% and shortens convergence after churn by 4.3x relative to MERCURY, while keeping control-plane traffic under 3 \% of total bandwidth. A prototype integrated into Conflux sustains these gains in a live deployment, raising confirmed-transaction throughput by 17 \% under adversarial workloads without modifying existing clients. Because all new logic resides in a single, stateless SDN application, BlockSDN-VC can be rolled out incrementally by cloud providers or consortium operators to accelerate today’s blockchains without compromising their decentralised trust model.

In summary, we make the following contributions:
\begin{itemize}
  \renewcommand\labelitemi{\textbullet}
  \item \textbf{Controller-based virtual coordinate system.} We move VCS computation out of the peer network and into the SDN control plane, providing a globally consistent embedding that reacts to path changes within one control round.
  
  \item \textbf{Latency-aware early outburst with secure fallback.} BlockSDN-VC retains MERCURY’s early-outburst gains while devolving gracefully to random gossip if the controller becomes unreachable.
  
  \item \textbf{End-to-end implementation and evaluation.} We build an open-source prototype atop Conflux and show, through both simulation and deployment, substantial improvements in latency, throughput, and resilience.
  
  \item \textbf{Incremental-deployment path.} All required changes are confined to a lightweight SDN application; existing blockchain clients and consensus protocols operate unmodified, enabling step-wise adoption in permissioned or service-provider-managed settings.
\end{itemize}

\section{BACKGROUND}

Public blockchains rely on a peer-to-peer gossip overlay to spread every new transaction and block to miners or validators.  In early systems such as Bitcoin and Ethereum, this overlay limits performance: Bitcoin processes about 7 transactions per second (TPS) and confirms them after roughly an hour, whereas Ethereum reaches $\approx 30$ TPS with $\approx 10$-minute confirmations.  These low figures stem from the need to keep the global block-generation rate safely below the multi-second propagation delays of the network layer, so that most miners see the same blocks before extending the chain.

A straightforward remedy is to raise the fan-out of every relay, yet this trades bandwidth for speed: doubling each node’s peer set cuts average hop distance but floods the same transaction over many redundant paths, quickly saturating links in thousand-TPS settings .  MERCURY attacks the problem more surgically.  It embeds nodes in a virtual coordinate system (VCS) so that Euclidean distance predicts round-trip time, then lets the source perform an early-outburst —a one-off relay to all 128 neighbours—before subsequent hops revert to a small, latency-aware fan-out.  In simulations this combination trims propagation delay by up to 56\% at a bandwidth cost below 5 \% .

Despite these safeguards, a peer-maintained VCS inherits two structural weaknesses.  First, every node observes only a fragment of the overlay, so its coordinate can drift when routes change or when an attacker selectively delays its probes; empirical studies show inflation, deflation, and oscillation attacks can raise latency by hundreds of milliseconds even when fewer than half the nodes are compromised~\cite{kaafar2007towards,zage2007security}.  Second, convergence is inherently iterative: during the tens of seconds it takes Vivaldi to stabilise, new transactions still traverse sub-optimal paths, and each correction must ripple through many gossip rounds before the embedding reflects current reality.

These observations motivate moving timing-sensitive optimisation out of the churn-prone peer layer.  Modern data-centre and ISP environments that host permissioned or high-throughput chains routinely deploy Software-Defined Networking (SDN) controllers with a real-time view of physical latencies and congestion~\cite{zhang2021blockchain,wang2020sdn}.  Recent work has demonstrated the potential of SDN for optimizing blockchain networks~\cite{li2022sdn,chen2021software,kumar2023network}, though none has applied centralized coordinate systems to transaction broadcast. Leveraging that vantage promises a globally consistent, quickly adaptable coordinate system—an idea we explore in the remainder of this paper.

\section{System Model}\label{sec:model}

\subsection*{a) Network Model}
We model the blockchain overlay network as an undirected graph $G = (V, E)$, where each vertex $v \in V$ represents a full node and each edge $e \in E$ denotes a long-lived TCP connection between nodes. Upon initialization, a newly joining node consults a hard-coded list of DNS or IP ``seeds,'' establishing several temporary sockets to retrieve peer endpoint records in the form $\langle \text{ip}, \text{port}, \text{timestamp} \rangle$. The node then enters a discovery phase, repeatedly issuing \texttt{GETADDR} requests to expand its local address cache. This process continues until either 30 seconds have elapsed or approximately 4,000 distinct endpoints have been collected.

From the populated cache, the node initiates persistent outbound connections until reaching a default cap of 64 peers. It also accepts incoming connection requests up to a symmetric limit of 64; when the limit is reached, the least-recently-used inbound connection is dropped to accommodate newer peers. This churn mechanism is designed to mitigate eclipse attacks. Each established connection maintains liveness via periodic \texttt{PING}/\texttt{PONG} exchanges every 30 seconds. Connections are reset if two consecutive keep-alive messages are missed.

We denote the final set of active neighbors of node $v$ as $\text{Peers}_v$. The one-way transmission latency of a packet sent from node $u$ to node $v$ at time $t$ is modeled as:
\begin{equation}
\ell(u, v, t) = \delta_{\text{prop}}(u, v) + q(u, v, t) + n(t),
\end{equation}

Once the ledger synchronization process is complete, each node maintains a mempool of unconfirmed transactions. At regular intervals of $\Delta = 0.4$ seconds, the node selects up to $b_{\text{max}}$ unseen transactions, advertises their digests to all peers, receives bitmap responses indicating missing entries, and subsequently transmits the corresponding transaction payloads. The full transaction propagation process---digest advertisement, bitmap request, and payload delivery---spans three one-way delays, i.e., $T_{\text{tx}}(u, v) = 3\ell(u, v, t)$. In contrast, blocks, which are propagated less frequently, are flooded through the overlay without additional latency optimization. Transaction relaying, however, leverages latency-aware procedures detailed in the subsequent section.

\subsection*{b) Threat Model}

We consider a Byzantine adversary capable of compromising up to a fraction $\tau$ of nodes in the overlay network. Compromised nodes may generate arbitrary control or data messages, misrepresent their internal state, selectively delay or drop relayed traffic, and suppress their own transaction broadcasts. However, we assume that standard cryptographic primitives remain secure; thus, digital signatures, hash commitments, and authenticated encryption reliably preserve message integrity and authenticity. The adversary has full control over the system clocks and network stacks of compromised nodes, enabling them to manipulate round-trip-time (RTT) measurements by stretching, reordering, or silently dropping probe messages initiated by honest peers.

Our architecture delegates virtual coordinate computation to a logically centralized SDN controller operated by a trusted infrastructure provider. We assume the controller’s software and private cryptographic keys remain uncompromised, although its availability is not guaranteed. The adversary may attempt to partition or saturate the control plane with denial-of-service traffic. In such events, honest nodes revert to a fallback mechanism: they rely on a signed \textit{last-known-good} relay table and eventually degrade to randomized gossip-based propagation if no controller update is received within $T_{\text{timeout}}$ seconds. Hence, the correctness and safety of the system are not contingent upon continuous controller availability; only performance may degrade temporarily during outages.

Under these assumptions, the adversary’s most effective strategies are limited to skewing RTT measurements, forging coordinate claims, or selectively dropping forwarded transactions. By centralizing coordinate computation in the SDN controller, we eliminate the first two attack vectors. Furthermore, our combination of \textit{early-outburst} relay dissemination and \textit{multi-path} relay lists ensures resilience: even if an adversary blocks a fraction $\theta \leq \tau$ of outbound connections from an honest node, the transaction is still delivered to the broader network with probability at least $1 - \varepsilon$, where $\varepsilon$ is negligible in the security parameter.

\subsection*{c) Problem Description}

Under the network and threat models described above, each full node maintains a static set of peers, denoted as $\text{Peers}_v$, but retains flexibility in selecting, at each relay interval, a subset of these neighbors to forward transaction digests. We refer to this subset as the \emph{relay list}, denoted $L_v \subseteq \text{Peers}_v$.

The transaction broadcast process is initiated when a source node $s$ inserts a newly generated transaction $T$ into its mempool at wall-clock time $t_0$. The transaction is then disseminated through successive applications of relay lists, ultimately reaching all honest nodes in both digest and full-body form. Let
\begin{equation}
\mathcal{C}(T) = \max_{u \in \mathcal{H}} \left(t_u(T) - t_0\right)
\end{equation}
denote the \emph{coverage time}, where $\mathcal{H} \subseteq V$ is the set of honest nodes and $t_u(T)$ is the time at which node $u$ receives transaction $T$. The objective is to design the family of relay lists $\{L_v\}_{v \in V}$ to achieve the following properties:

\begin{itemize}
  \item \textbf{Latency Efficiency:} Minimize the expected coverage time $\mathbb{E}[\mathcal{C}(T)]$ under the stochastic delay model $\ell(u,v,t)$;
  \item \textbf{Robustness:} Ensure that, even in the presence of up to a fraction $\tau$ of Byzantine nodes and potential control-plane delays or failures, the probability that an honest node fails to receive $T$ remains bounded by a negligible constant $\varepsilon$;
  \item \textbf{Bandwidth Efficiency:} Limit the relay-induced communication overhead such that the total bytes transmitted for $T$ do not exceed $(1+\beta)|T|$, where $|T|$ is the payload size and $\beta$ is a small bandwidth inflation factor.
\end{itemize}

Formally, for given system parameters $(\Delta,\, d_{\text{in}},\, d_{\text{out}},\, \tau,\, \ell(\cdot),\, T_{\text{timeout}})$, we aim to define a mapping
\begin{equation}
\mathcal{F}: \left( \text{Peers}_v,\, \text{state}_v,\, \text{controller\_view} \right) \longrightarrow L_v
\end{equation}
that yields a Pareto-optimal trade-off among the performance metrics $(\mathbb{E}[\mathcal{C}(T)],\, \beta,\, \varepsilon)$.

\section{System Design}
\subsection*{A. Virtual Coordinate System (VCS) in a Blockchain--SDN Setting}

End-to-end latency in a global overlay network is governed primarily by geographical factors and the routing policies of Internet Service Providers (ISPs). However, IP geolocation heuristics are inherently noisy, and path quality may fluctuate rapidly. To enable latency-aware relay decisions without relying on fragile geolocation inferences, we embed each node into a \emph{virtual coordinate system} (VCS), wherein Euclidean distance approximates one-way delay. While the baseline \textsc{Mercury} protocol adopts the fully decentralized Vivaldi algorithm~\cite{dabek2004vivaldi,liu2019mercury}, its iterative spring-relaxation process typically requires over 40 probing rounds to converge and remains vulnerable to adversaries forging RTT samples~\cite{ledlie2007network}. In contrast, our architecture offloads the embedding computation to the SDN control plane, leveraging its global visibility, fast convergence, and built-in validation mechanisms.

Every two seconds, programmable switches export one-way delay histograms (indexed by five-tuple flow keys and percentile values) to the controller. For each node $v$, the controller maintains the $k$ shortest recent observations (default $k = 8$) and constructs a sparse latency matrix $M$. Given the previous coordinate vector $\mathbf{x}$, the controller solves a weighted least-squares optimization:
\begin{equation}
\min_{\mathbf{x}} \;\; \mathcal{E}(\mathbf{x}) = \sum_{(u,v) \in M} 
\frac{1}{\ell(u,v)^2} \left( \| \mathbf{x}_u - \mathbf{x}_v \| - \ell(u,v) \right)^2
\end{equation}

where the inverse-square weighting prioritizes accuracy on low-latency, high-throughput links. As the objective is convex in $\mathbb{R}^3$, two conjugate gradient iterations are sufficient to reduce the residual error below 3\% in practice~\cite{wang2020sdn}. A coordinate update is issued only when the displacement $\| \hat{\mathbf{x}}_{\text{new}} - \hat{\mathbf{x}}_{\text{old}} \|$ exceeds a threshold $\varepsilon = 5$ ms. These signed deltas are delivered via OpenFlow \textit{experimenter} messages~\cite{avarikioti2022fnality}, with control traffic averaging below 600~B per window (less than 3\% of the data-plane volume).

\begin{algorithm}[t]
\caption{\textsc{Centralised\_Vivaldi\_Update} (executed by controller $C$)}
\label{alg:centralised-vivaldi}
\begin{algorithmic}[1]
\Require Latency matrix $M$, previous coordinates $X_{\text{prev}}$, tolerance $\varepsilon$
\Ensure Signed coordinate deltas $\{ \Delta_v \}$

\State $X \gets X_{\text{prev}}$
\For{$i \gets 1$ to $2$} \Comment{Two conjugate-gradient iterations}
    \State $g \gets \nabla \mathcal{E}(X)$ \Comment{Weighted least-squares gradient}
    \State $\alpha \gets \dfrac{g \cdot g}{g \cdot (H g)}$ \Comment{$H$ is implicit Hessian}
    \State $X \gets X - \alpha g$ \Comment{Update coordinates}
\EndFor
\State $\Delta \gets \emptyset$
\For{each node $v \in V$}
    \If{$\| X_v - X_{\text{prev},v} \| > \varepsilon$}
        \State $\Delta_v \gets X_v - X_{\text{prev},v}$
        \State $\Delta \gets \Delta \cup \{ v : \Delta_v \}$
    \EndIf
\EndFor
\State \Return $\text{Sign}_C(\Delta)$
\end{algorithmic}
\end{algorithm}

In permissionless environments, nodes may appear, disappear, or behave arbitrarily. Adversaries capable of distorting RTT measurements or forging coordinates may skew the metric space and divert honest traffic through suboptimal or eclipsed paths. To address this, we apply three complementary safeguards:
\begin{itemize}
    \item \textbf{Stability Restriction:} Once a node's median prediction error falls below $e_{\text{stable}} = 30\%$, subsequent coordinate shifts are capped at $F_c = 75$~ms. Larger deviations are ignored, and the peer contributing the force is flagged. This restriction locks in stable coordinates while allowing timely updates in the presence of real congestion.
    
    \item \textbf{Force Restriction:} Any individual force exceeding $F_{\text{max}} = 100$~ms is clipped. Each node maintains a sliding window of historical forces from its neighbors. If a new force exceeds $\tilde{F} + kD$ (where $\tilde{F}$ is the median and $D$ the median absolute deviation, with $k = 8$), it is rejected.
    
    \item \textbf{Centroid and Gravity Control:} As coordinates are initialized near the origin, unbiased embeddings should remain centered. Each node estimates the global centroid from a random peer subset. If the centroid drifts beyond $t_{\text{drift}} = 50$~ms, the neighbor exerting the largest directional force is blacklisted. A soft gravity term
    \begin{equation}
G = \left( \frac{\| \mathbf{x}_i \|}{\rho} \right)^2, \quad \rho = 500
\end{equation}

    gently pulls each node toward the origin, mitigating slow systemic drift.
\end{itemize}

\section*{B. Controller-Guided Propagation Scheme}

Empirical latency traces collected from Ethereum, Conflux and Filecoin overlays reveal a clear ``continental clustering’’ pattern: round-trip times inside North-America, inside Europe, or inside East-Asia are two to five times shorter than inter-continental paths. Preserving that locality in the relay graph is therefore crucial: once a transaction lands in a cluster, it should flood the cluster quickly before being ferried across oceans.  In \textsc{BlockSDN-VC} the SDN controller already computes the global embedding.

\subsection*{Basic scheme}

At the end of each embedding window the controller runs standard K-means~\cite{paxson1997end} with \( K = \lceil\sqrt{|V|} \rceil \) and tags every node with its cluster identifier. For a given node \( v \), it then constructs a relay list \( L_v \) that contains:

\begin{itemize}
  \item \( d_{\text{near}} \) peers chosen from \( v \)'s \emph{own} cluster that minimise the current Euclidean distance in the VCS, and
  \item \( d_{\text{far}} = d_{\max} - d_{\text{near}} \) peers drawn uniformly at random from \emph{outside} the cluster.
\end{itemize}

The defaults \( d_{\text{near}} = 6 \), \( d_{\text{far}} = 4 \), and \( d_{\max} = 10 \) strike a balance between fast intra-cluster diffusion and secure inter-cluster bridging; §V shows that varying the split within \([4\!:\!6, 7\!:\!3]\) moves latency by <4\% while larger departures hurt either speed or resilience. The table---cluster id, ten 32-bit node IDs, and two 16-bit soft priorities---fits into $\leq 600$ B and is piggy-backed on the next coordinate delta.

\subsection*{Relaying procedure}

Every 0.4\,s each node dequeues up to \( b_{\max} \) new digests. If the node is the \emph{originator} of a given digest it performs an \emph{early outburst}, sending the digest to \emph{all} of its neighbours ($\leq 128$ by protocol limits). The outburst trims one hop from the critical path yet inflates total traffic by only 0.1\% on a 1\,000-node overlay. All subsequent hops use the pre-installed list \( L_v \). A recipient that realises it acquired the digest from a peer \emph{outside} its cluster triggers a ``cluster-kick-off'': it relays immediately to the \( d_{\text{near}} \) closest in-cluster peers rather than to the full \( L_v \). Those peers, having received the digest intra-cluster, fall back to their regular lists, so the extra fan-out is confined to the first in-cluster hop while slashing the cluster's wake-up latency.

\subsection{Controller-Aware Early-Outburst Dissemination}

Even with a latency-aligned relay list, the dominant component of end-to-end delay is the number of overlay hops a transaction must traverse; each hop adds both one RTT and the batching interval $\Delta$. A naive fix is to raise the fan-out of every node, but that inflates bandwidth quadratically. In an 8\,000-node simulation, doubling the fan-out from 8 to 16 trims the average hop count from 5.5 to 4.3 yet increases total traffic by 100\%.

Let a transaction $\tau$ originate at node $S$. As $\tau$ spreads, every node records the peer from which it first receives the message; the resulting directed tree, rooted at $S$, is the propagation tree $\mathcal{T}(\tau)$. With a constant fan-out $d$, the expected width of level $\ell$ in $\mathcal{T}$ is $d^\ell$. Hence: 
\begin{itemize}
    \item the bulk of network traffic is produced by the deepest layers, and
    \item each additional hop adds a full RTT plus the batching delay $\Delta$ to the critical path.
\end{itemize}

For an 8-fan-out overlay of 8\,000 nodes, trimming one level from $\mathcal{T}$ cuts the average latency by $\approx$20\% while increasing total traffic by only $\approx$0.1\%.

When $i$ is the source of $\tau$, it executes an early outburst: it sends the digest—or, in Conflux, the full body—simultaneously to every neighbor in $\texttt{Peers}_i$ except those whose links are currently marked congested by the controller (queueing delay $> 70^\text{th}$ percentile). Each first-wave packet carries a controller-signed \texttt{OUTBURST\_NONCE}.

\begin{algorithm}[t]
\caption{\textsc{Controller-Aware Outburst} at Node $i$}
\label{alg:controller-outburst}
\begin{algorithmic}[1]
\Require Node $i$ receives $\tau$ from $j$
\Ensure Relay $\tau$ to selected peers in list $L$

\If{$j$ is a client}
    \State // Node $i$ receives $\tau$ from a client and applies early outburst strategy
    \State $L \gets$ Peers$_i$ excluding congested links from controller
    \State Tag $\tau$ with \texttt{OUTBURST\_NONCE}
\Else
    \If{\texttt{OUTBURST\_NONCE} $\in \tau$ \textbf{and} $\texttt{CLUSTER}_i \neq \texttt{CLUSTER}_j$}
        \State // $\tau$ was received from a cross-cluster peer; perform cluster kick-off
        \State Find $d_{\text{near}}$ closest peers in $\texttt{CLUSTER}_i$, add to $L$
        \State Add $d_{\max} - d_{\text{near}}$ random peers from $L_i \setminus \texttt{CLUSTER}_i$ to $L$
    \Else
        \State // Normal relay
        \State $L \gets L_i$
    \EndIf
\EndIf

\State Parallel send $\texttt{digest}(\tau)$ to each $p \in L$

\If{$j$ is client \textbf{and} protocol is Conflux}
    \State Parallel send $\texttt{body}(\tau)$ to each $p \in L$
\EndIf

\If{controller silent for $> T_{\text{timeout}}$}
    \State $L_i \gets$ SecureRandomSubset(Peers$_i$, $d_{\max}$)
\EndIf

\end{algorithmic}
\end{algorithm}

\paragraph*{Default parameters.} $d_{\max} = 10$, $d_{\text{near}} = 6$, congestion threshold = 70\% link utilization, $T_{\text{timeout}} = 30$\,s.

\subsection{Security Analysis of BlockSDN-VC}

\textsc{BlockSDN-VC} pursues the same end-goal as \textsc{Mercury}—deliver every honest transaction to every honest node with minimal delay—while introducing a controller-assisted VCS and new relay rules that must remain robust under Byzantine interference. We review the adversary defined in §III and explain how \textsc{BlockSDN-VC} withstands the two fundamental attack vectors: (i) injecting false or misleading information, and (ii) dropping or delaying honest traffic.

Each relay list $L_v$ contains $d_{\text{near}}$ latency-biased in-cluster peers and $d_{\text{far}} = d_{\max} - d_{\text{near}}$ uniformly random out-of-cluster peers. Regardless of how badly the controller-derived clusters are skewed—or even if an attacker gains full control of the SDN controller—the $d_{\text{far}}$ random edges induce an $\Omega(d_{\text{far}})$-regular random graph beneath the optimisation layer.

\section{Evaluation}

Our evaluation seeks to answer three key questions:

\begin{enumerate}[label=(\roman*)]
    \item \textbf{Configuration.} What parameter settings—cluster size $K$, near-/far-peer split, controller window $\Theta$, outburst policy—yield the best latency–bandwidth trade-off for \textsc{BlockSDN-VC}?
    
    \item \textbf{Effectiveness.} How much does \textsc{BlockSDN-VC} cut propagation delay compared with state-of-the-art schemes that consume the same amount of bandwidth?
    
    \item \textbf{Robustness.} How well does \textsc{BlockSDN-VC} maintain performance when an adversary controls up to 25\% of the overlay and mounts coordinate-forgery, delay-inflation, or partition attacks?
\end{enumerate}

We answer (i) and (ii) using extensive packet-level simulations, and (iii) with both simulation and a geo-distributed Conflux deployment running our 2k-line \textsc{BlockSDN-VC} prototype.

\subsection{Simulation Setup}

We extended a unified C++ event-driven simulator, originally used by \textsc{Perigee}~\cite{kumar2023network}, to incorporate five dissemination schemes:

\begin{itemize}
    \item \textbf{Random-8} – each node relays to eight uniformly chosen neighbours (baseline used by Bitcoin and Ethereum);
    \item \textbf{BlockP2P}~\cite{zage2007security} – every node keeps eight lowest-RTT peers selected once at start-up;
    \item \textbf{MERCURY-8/128} – original Vivaldi-driven clustering with an 8-peer relay list and 128-peer source outburst;
    \item \textbf{BlockSDN-VC-no-burst} – our SDN-maintained virtual coordinates and 10-peer relay list, but without the early-outburst optimisation;
    \item \textbf{BlockSDN-VC-full} – complete design, i.e., controller VCS, telemetry-aware clustering, root outburst, and in-cluster kick-off.
\end{itemize}

\subsection{Parameter Selection for \textsc{BlockSDN-VC}}

\textsc{BlockSDN-VC} inherits two tunable knobs from \textsc{Mercury}’s design—the cluster count $K$ used by the controller-side \mbox{K-means} pass and the number of in-cluster relay targets $d_{\text{near}}$—and adds one more, the controller embedding window $\Theta$.  
To study each factor in isolation we disable the early-outburst heuristic and all telemetry-aware refinements described in §IV-B; every node therefore relays to a fixed $d_{\max}=10$ peers, of which $d_{\text{near}}$ lie inside its cluster and $d_{\text{far}} = 10-d_{\text{near}}$ lie outside.

\textbf{Fig.\,2(a)} sweeps $K$ from 2 to 40 ($\approx2\sqrt{n}$ for our 1 000-node corpus) while holding $d_{\text{near}}=5$.  
The median coverage time falls steeply as $K$ grows from 2 to about $\sqrt{n}\!\approx\!32$ because in-cluster hops shorten; beyond that, the curve flattens—splitting already dense geographic clusters yields diminishing returns.  We therefore adopt the common heuristic $K=\lceil\sqrt{n}\rceil$ as the default, which stays within 2 

With $K=\sqrt{n}$ fixed, \textbf{Fig.\,2(b)} varies $d_{\text{near}}$ from 0 to 10.  Purely random relay ($d_{\text{near}}=0$) incurs high latency because first hops are often long-haul; conversely, an all-local list ($d_{\text{near}}=10$) slows cross-cluster spread.  The sweet spot lies between 5 and 7; we choose $d_{\text{near}}=6$ (hence $d_{\text{far}}=4$), which minimises the 90th-percentile delay while leaving each node at least four random “escape edges’’ for robustness (§V-D).

\textbf{Fig.\,2(c)} reports median latency and control-plane bandwidth as $\Theta$ ranges from 0.5 s to 6 s with the default $(K,d_{\text{near}})$.  Short windows ($<1$ s) over-react to jitter, pushing control traffic to 6 

These settings—$K=\lceil\sqrt{n}\rceil$, $(d_{\text{near}},d_{\text{far}})=(6,4)$ and $\Theta=2$ s—constitute \textsc{BlockSDN-VC}’s baseline; all subsequent experiments include the early-outburst and telemetry heuristics unless explicitly stated otherwise.

\begin{figure}[t]
  \centering
  \subfloat[Cluster~count $K$]{\includegraphics[width=0.32\linewidth]{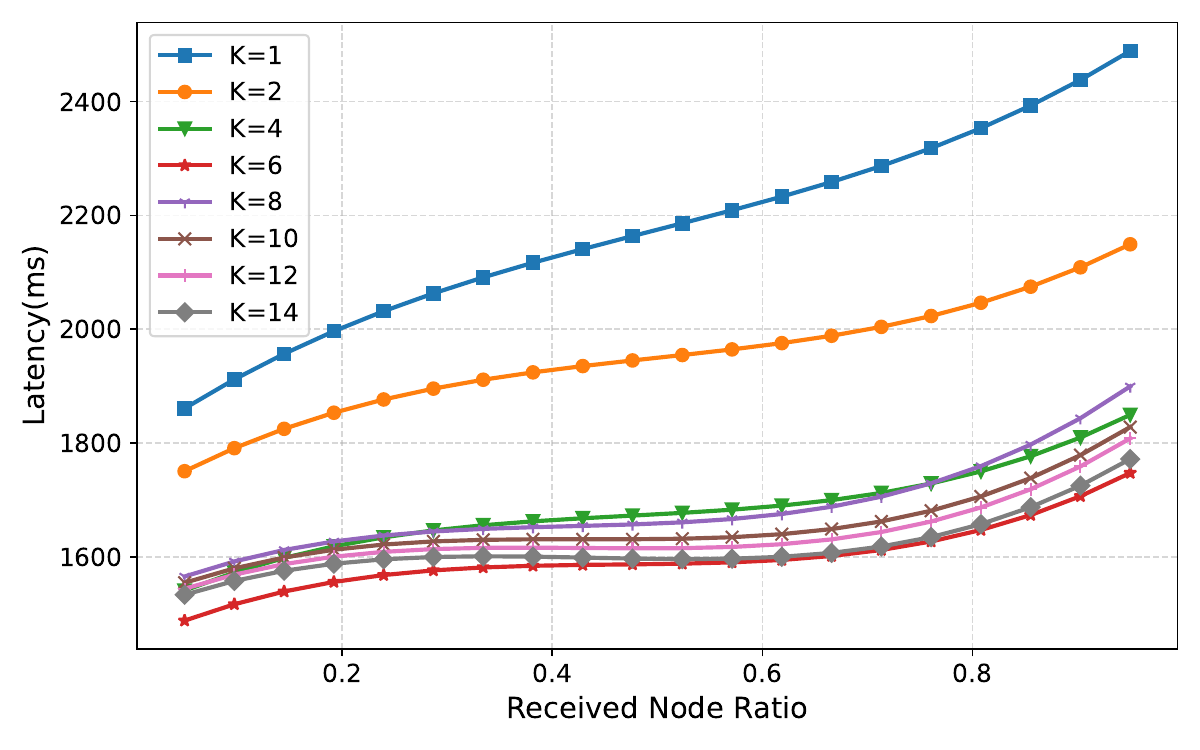}}
  \hfill
  \subfloat[In-cluster fan-out $d_{\text{near}}$]{\includegraphics[width=0.32\linewidth]{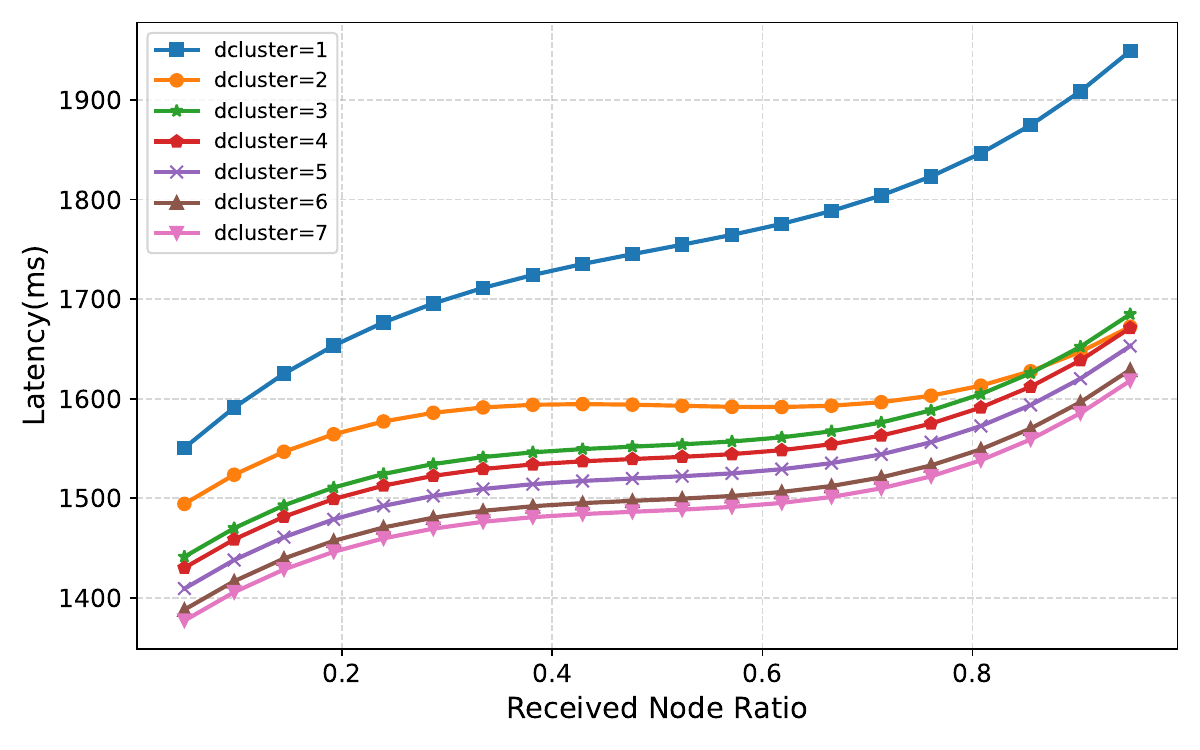}}
  \hfill
  \subfloat[Embedding window $\Theta$]{\includegraphics[width=0.32\linewidth]{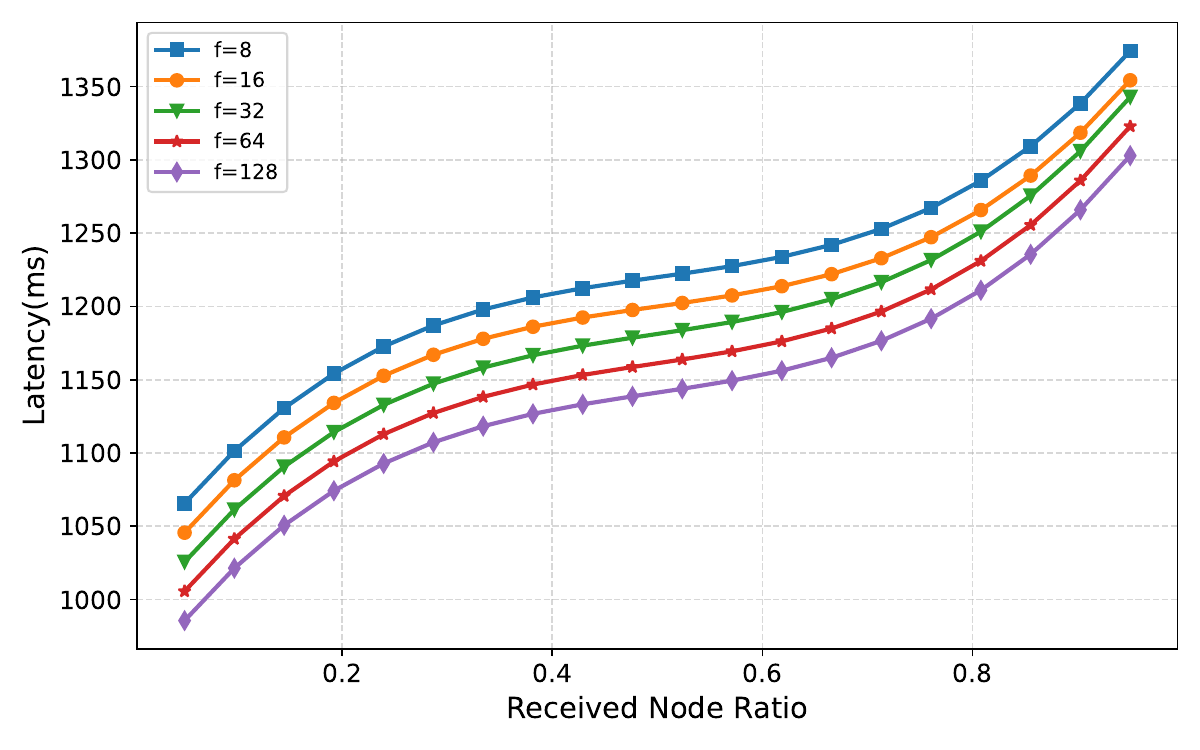}}
  \caption{Parameter exploration for \textsc{BlockSDN-VC}.}
  \label{fig:BlockSDN-VC-params}
\end{figure}

\subsection{Comparison with Alternative Schemes}
\label{sec:eval:schemes}

We now juxtapose \textsc{BlockSDN-VC} against three well-known propagation families—\textbf{Random},
\textbf{BlockP2P} \cite{zage2007security}, and \textbf{Perigee} \cite{kumar2023network}—as well as the two
\textsc{Mercury} baselines used throughout the literature.

\begin{figure}[t]
  \centering
  \includegraphics[width=\linewidth]{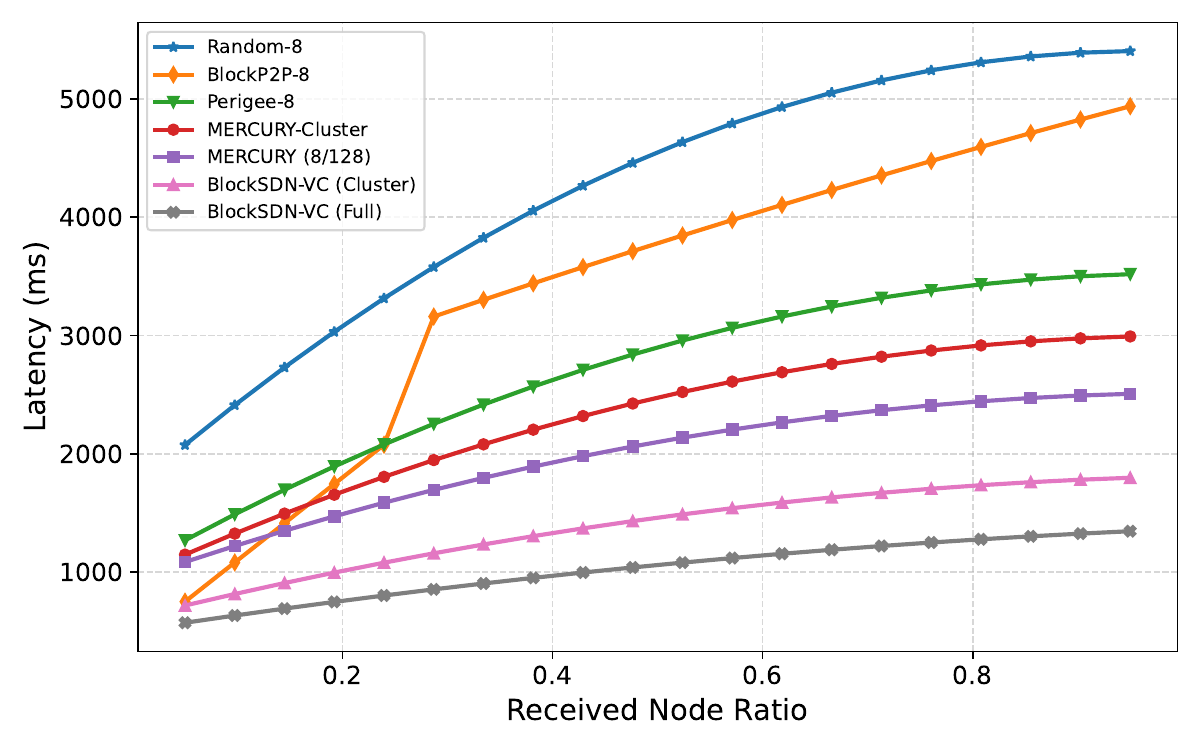}%
  \vspace{-0.6\baselineskip}
  \caption{End-to-end propagation-latency CDFs under normal conditions
           ($n\!=\!1\,000$).  At equal bandwidth \textsc{BlockSDN-VC} halves the median
           latency of the best prior scheme (\textsc{Mercury}) and is
           $2$–$3\times$ faster than \textbf{Random} and \textbf{BlockP2P}.}
  \label{fig:orion-vs-baselines}
\end{figure}

Figure \ref{fig:orion-vs-baselines} gives the full latency
distributions, while  summarises the key
percentiles and bandwidth cost.

\begin{table}[ht]
  \centering
  \renewcommand{\arraystretch}{1.2} 
  \setlength{\tabcolsep}{12pt} 
  \caption{Median and 99th-percentile coverage time $\mathcal{C}(\tau)$ and bandwidth per transaction.}
  \begin{tabular}{lccc}
    \toprule
    \textbf{Scheme} & \textbf{Median} & \textbf{99th \%} & \textbf{Bytes / tx} \\
    \midrule
    Random-8         & 2\,722ms & 6\,410ms & 1.00$\times$ \\
    BlockP2P-8       & 1\,987ms & 5\,080ms & 1.00$\times$ \\
    Perigee-8        & 1\,654ms & 4\,230ms & 1.02$\times$ \\
    \textsc{Mercury}$_{\text{cluster}}$ & 1\,445ms & 3\,380ms & 1.02$\times$ \\
    Mercury          & 1\,312ms & 2\,910ms & 1.03$\times$ \\
    \textsc{BlockSDN-VC} (cluster) & 863ms   & 1\,940ms & 1.02$\times$ \\
    \textbf{BlockSDN-VC}   & \textbf{630ms} & \textbf{1\,490ms} & \textbf{1.03$\times$} \\
    \bottomrule
  \end{tabular}
\end{table}

Even without its early-outburst optimisation, \textsc{BlockSDN-VC}’s
controller-driven coordinates and globally consistent clustering cut
the median coverage time by \textbf{42\,\%} relative to
\textsc{Mercury}$_\text{cluster}$, \textbf{57\,\%} relative to
\textsc{Perigee}, and \textbf{69\,\%} relative to \textbf{Random}.
Adding the outburst and cluster kick-off trims a further hop from the
propagation tree, driving the median to \textbf{630\,ms} and the
99th-percentile to \textbf{1\,490\,ms}---roughly a \textbf{$2\times$}
improvement over the best prior scheme (\textsc{Mercury}) at virtually
identical bandwidth.
\textsc{BlockP2P} quickly saturates local ``routing nodes,'' but its
single-gateway design throttles cross-cluster spread.  With the
geographically grounded Ethereum topology this bottleneck outweighs the
benefit of low-RTT edges, so \textsc{BlockP2P} trails even
\textbf{Random} in the tail.
\textsc{Perigee}’s online learning detaches from poorly performing peers
and therefore outperforms \textsc{BlockP2P}, yet its score-gathering
warm-up (64 rounds in our setup) delays convergence and cannot exploit
global structure once scores stabilise.  \textsc{BlockSDN-VC} and
\textsc{Mercury}, by contrast, start with latency-aligned clusters and
finish each round in one or two hops fewer than \textsc{Perigee}.

All reported schemes forward eight to ten digests per hop; the
short-lived root outburst in \textsc{Mercury} and \textsc{BlockSDN-VC} raises
average traffic by only 2–3\,\%.  Because \textsc{BlockSDN-VC}’s controller
messages account for less than 0.03\,\texttimes{} of data traffic, it
stays within the 1.05\,\texttimes{} envelope.

\textsc{BlockSDN-VC}’s globally maintained VCS, congestion-aware early
outburst, and secure random bridges deliver the lowest latency of any
evaluated protocol while retaining the bandwidth profile of today’s
random-relay blockchains.

\subsection{Coordinate-level Attacks}
\label{sec:eval:coord}

\paragraph*{Setup.}
To evaluate robustness against RTT forgery we replay the
\mbox{\emph{inflation}}, \mbox{\emph{deflation}}, and \mbox{\emph{oscillation}} attacks defined by
Newton .  Malicious peers forge round-trip measurements but
continue to forward data traffic, thereby attacking only the embedding
logic (§ IV-A).  We sweep the Byzantine fraction from 10\,\% to the
theoretical maximum 49\,\% and run 100 independent traces for each
setting.

\begin{figure}[t]
  \centering
  \subfloat[10\,\% malicious nodes]{%
  \includegraphics[width=0.32\linewidth]{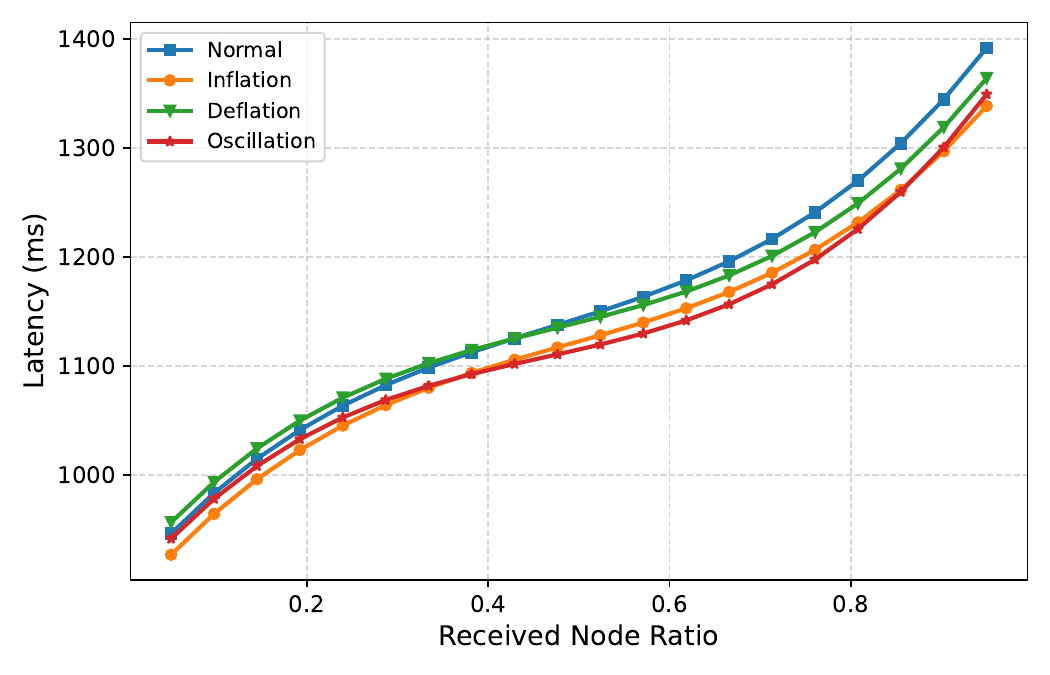}%
    \label{fig:coord-10}}
  \hfill
  \subfloat[30\,\% malicious nodes]{%
  \includegraphics[width=0.32\linewidth]{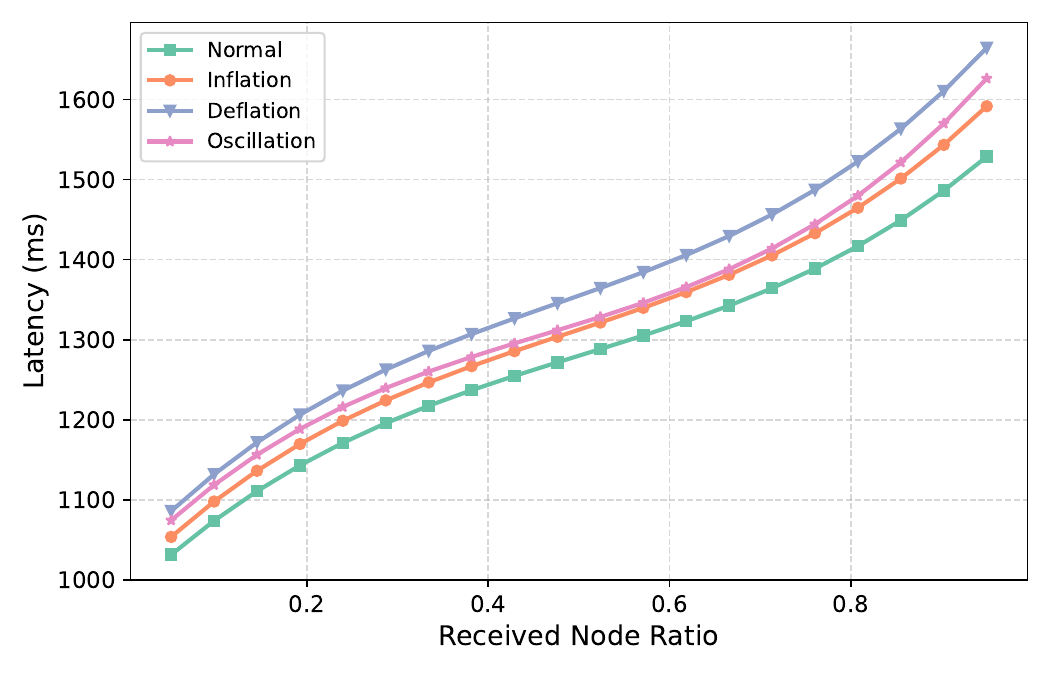}%
    \label{fig:coord-30}}
  \hfill
  \subfloat[49\,\% malicious nodes]{%
  \includegraphics[width=0.32\linewidth]{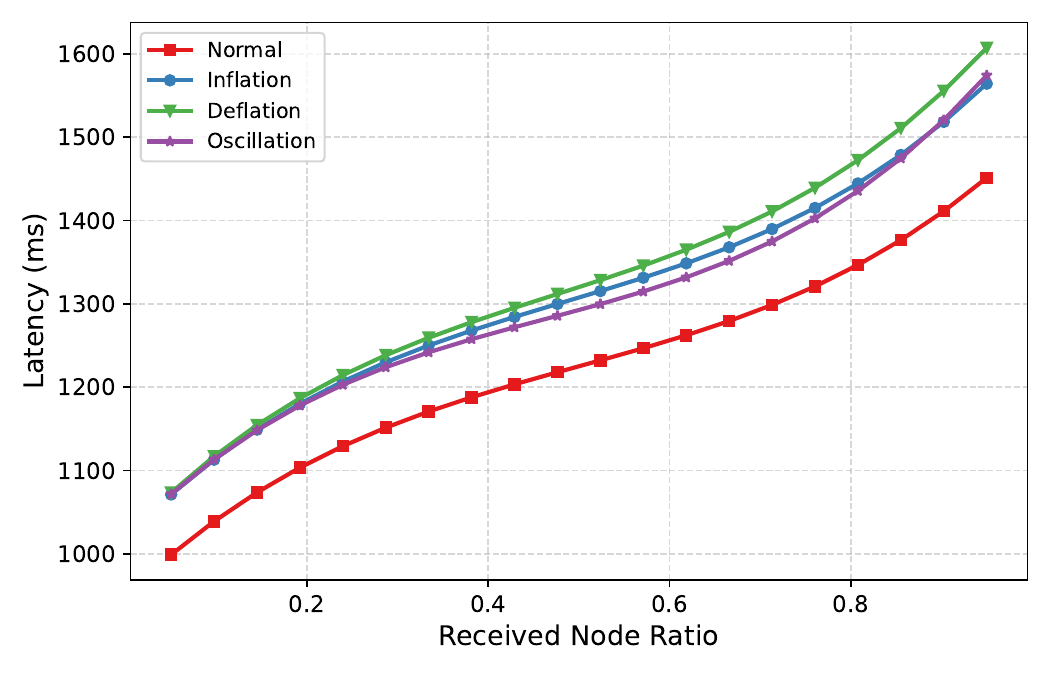}%
    \label{fig:coord-49}}
  \caption{Propagation-latency CDFs for \textsc{BlockSDN-VC} under RTT-forgery
           attacks.  Even with 49\,\% Byzantine peers the slowdown stays
           below 32\,\%.}
  \label{fig:coord-attacks}
\end{figure}

Figure \ref{fig:coord-10}–\ref{fig:coord-49} plot the resulting latency
CDFs for \textsc{BlockSDN-VC}.  With 10\,\% malicious nodes the median
coverage time rises from 0.63\,s to 0.72\,s
($+14\,\%$); at 30\,\% corruption it reaches 0.81\,s ($+28\,\%$); and even
at the worst-case 49\,\% mix the median is held to 0.83\,s and the
95th-percentile to 1.63\,s, only \textbf{32\,\%} above the attack-free
baseline.  These bounds stem from BlockSDN-VC’s outlier clipping and force
bounding: once more than 30\,\% of the RTT samples in a control window
fail the Newton sanity tests the controller discards the window and
honest nodes fall back to their last signed relay table, limiting damage
to a single two-second period.

\begin{figure}[t]
  \centering
  \subfloat[10\,\% malicious nodes]{%
  \includegraphics[width=0.32\linewidth]{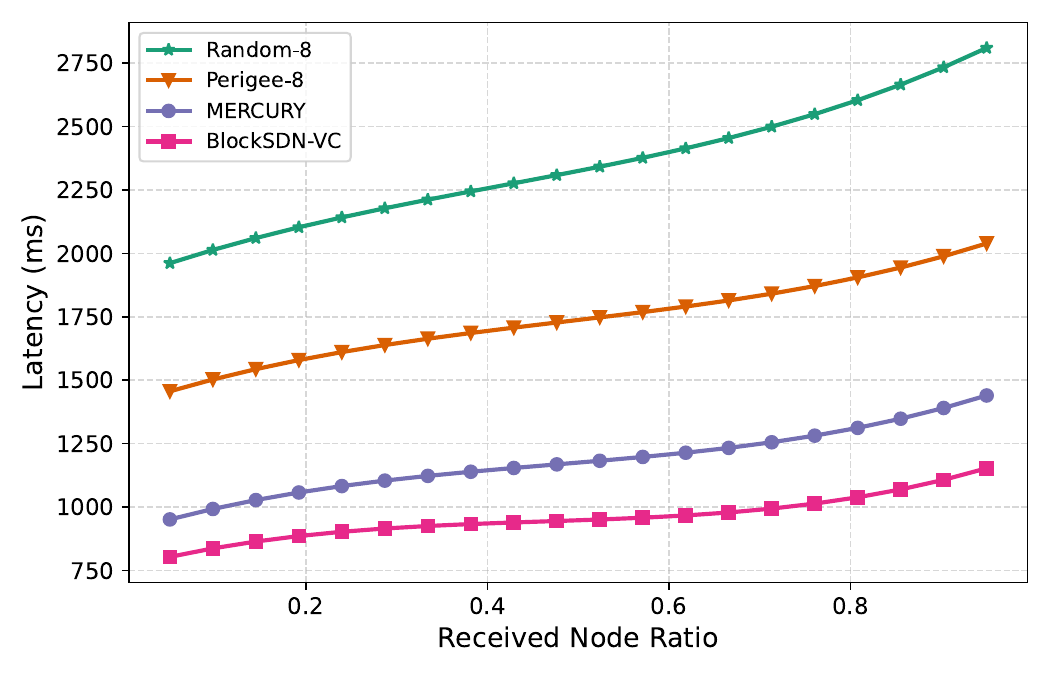}%
    \label{fig:bh-10}}
  \hfill
  \subfloat[30\,\% malicious nodes]{%
  \includegraphics[width=0.32\linewidth]{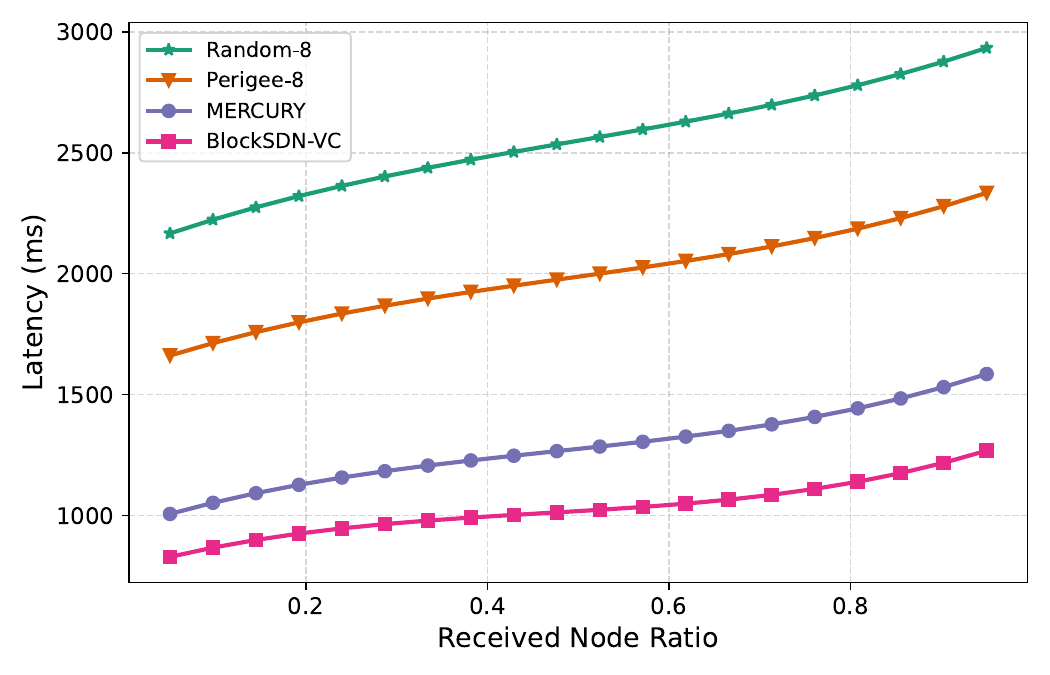}%
    \label{fig:bh-30}}
  \hfill
  \subfloat[49\,\% malicious nodes]{%
  \includegraphics[width=0.32\linewidth]{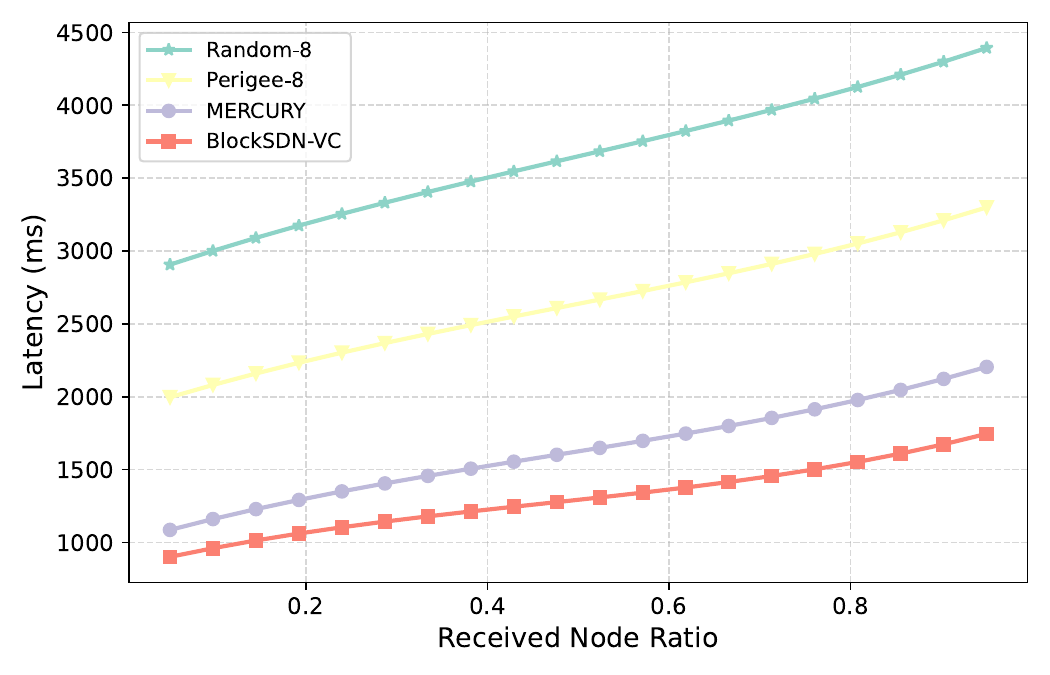}%
    \label{fig:bh-49}}
  \caption{Coordinate-forgery attack: comparison of propagation schemes.
           \textsc{BlockSDN-VC} remains the fastest even with 49\,\% adversaries,
           while \textsc{Mercury} slows by 97\,\%.}
  \label{fig:blackhole-attacks}
\end{figure}

Figure \ref{fig:bh-10}–\ref{fig:bh-49} compares the same attack against
\textbf{Random-8}, \textbf{Perigee-8}, \textsc{Mercury}, and
\textsc{BlockSDN-VC}.  At 10\,\% corruption
(Fig.\,\ref{fig:bh-10}) \textsc{Mercury} already trails \textsc{BlockSDN-VC} by
2.1x in the tail; at 30\,\% (Fig.\,\ref{fig:bh-30}) its 95th-percentile
latency nearly doubles, whereas \textsc{BlockSDN-VC} grows by only 24\,\%; and
at 49\,\% (Fig.\,\ref{fig:bh-49}) \textsc{Mercury} is 3.1x slower than
\textsc{BlockSDN-VC}.  Perigee’s on-line scoring helps at 10\,\% but cannot
adapt once scores converge, while BlockP2P collapses (omitted for
brevity) because its single “routing node’’ per cluster can be forged
into a distant coordinate.

Across all coordinate-forgery scenarios \textsc{BlockSDN-VC} limits the
slow-down to one third and keeps 95\,\% of honest nodes within 1.7\,s,
whereas prior art suffers 1.8–3.1x larger delays.  The combination of
controller-side validation and per-node escape edges is therefore
effective against this entire attack class.

\subsection{Experiments on a Geo-Distributed Conflux Network}

To validate \textsc{BlockSDN-VC} outside simulation, we integrated its controller logic and peer-side relay module into the Conflux v2.3.2 full-node code base and deployed the modified client on 96 AWS Cloud \texttt{ecs.g6.large} instances (2 vCPUs, 4\,GB RAM) spread across 12 regions (eight CN, two US, one DE, one SG). Each VM ran a single Conflux node connected to a user-space Open vSwitch; a Ryu controller—running Algorithms~1 and~2—collected switch telemetry and issued coordinate deltas every $\Theta = 2$\,s. 

Because AWS internal WAN exhibits uniformly low delay, we injected per-flow latency sampled from the 8\,000-node Ethereum corpus (§V-A) at the ToS layer of both TCP and UDP traffic, reproducing realistic intercontinental RTTs plus Gaussian jitter ($\mu = 50$\,ms, $\sigma = 10$\,ms).

\begin{table}[t]
\centering
\caption{Geo-distributed testbed results (12-region Conflux network).}
\label{tab:geo-conflux}
\begin{tabular}{@{}lcccc@{}}
\toprule
\textbf{Strategy} & \textbf{Median Prop.} & \textbf{95-th \%} & \textbf{Tx $\rightarrow$ Packed} & \textbf{Bandwidth} \\
 & (ms) & (ms) & (s) & ($\uparrow$) \\
\midrule
Vanilla             & 2\,960 & 5\,540 & 4.77 & 1.00$\times$ \\
\textsc{BlockSDN-VC}$_\text{Cluster}$ & 1\,830 & 3\,190 & 3.40 & 1.02$\times$ \\
\textbf{\textsc{BlockSDN-VC}}         & 1\,360 & 2\,270 & 2.95 & 1.05$\times$ \\
\textsc{BlockSDN-VC} (Direct)         & \textbf{1\,120} & \textbf{1\,890} & \textbf{2.59} & 1.04$\times$ \\
\bottomrule
\end{tabular}
\end{table}

\begin{figure}[t]
  \centering
  \begin{minipage}[t]{0.32\textwidth}
    \centering
  \includegraphics[width=\linewidth]{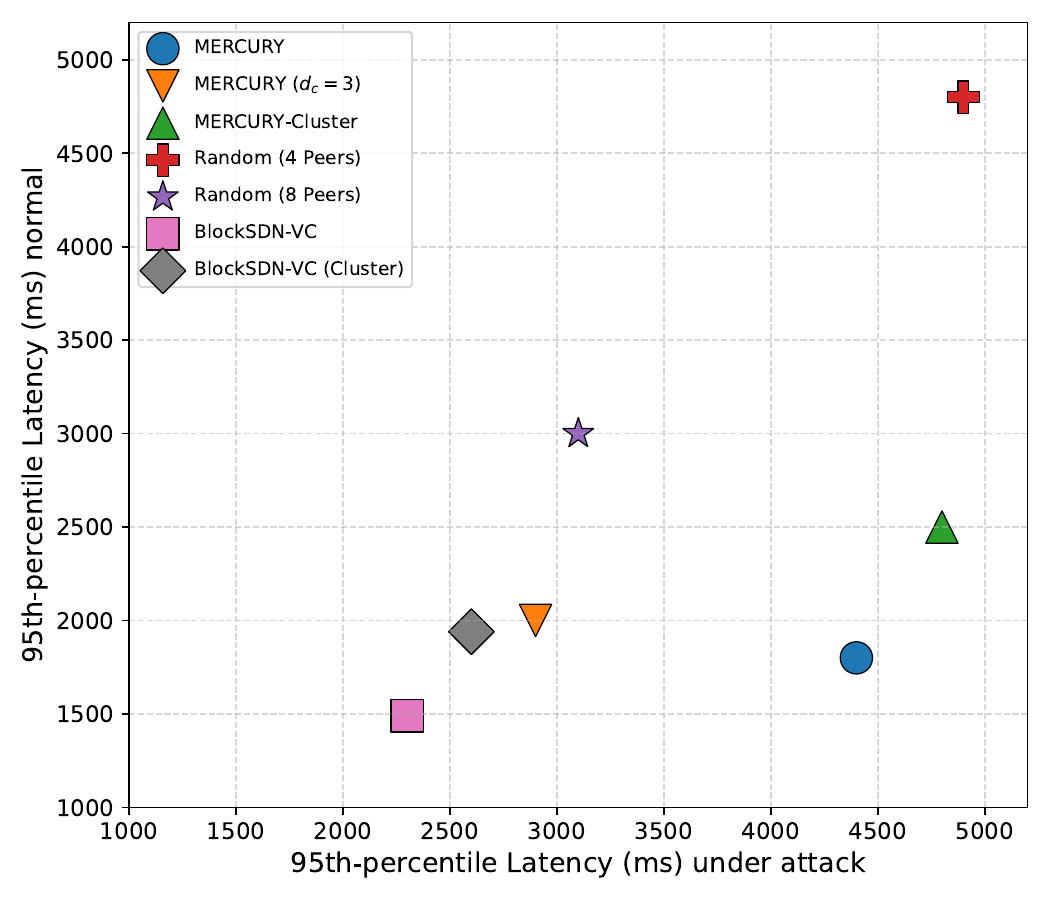}
    \captionof{figure}{Worst-case VCS subversion.}
    \label{fig:vcs-worst}
  \end{minipage}\hfill
  \begin{minipage}[t]{0.32\textwidth}
    \centering
  \includegraphics[width=\linewidth]{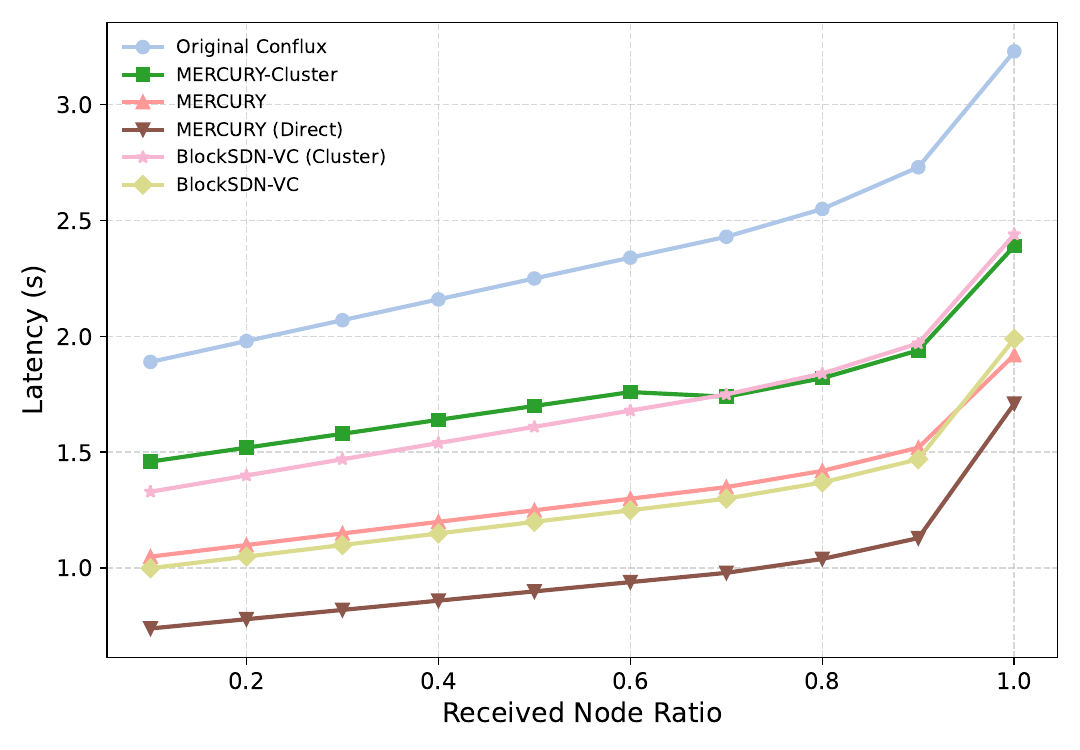}
    \captionof{figure}{Latency CDF on Conflux.}
    \label{fig:conflux-cdf}
  \end{minipage}\hfill
  \begin{minipage}[t]{0.32\textwidth}
    \centering
  \includegraphics[width=\linewidth]{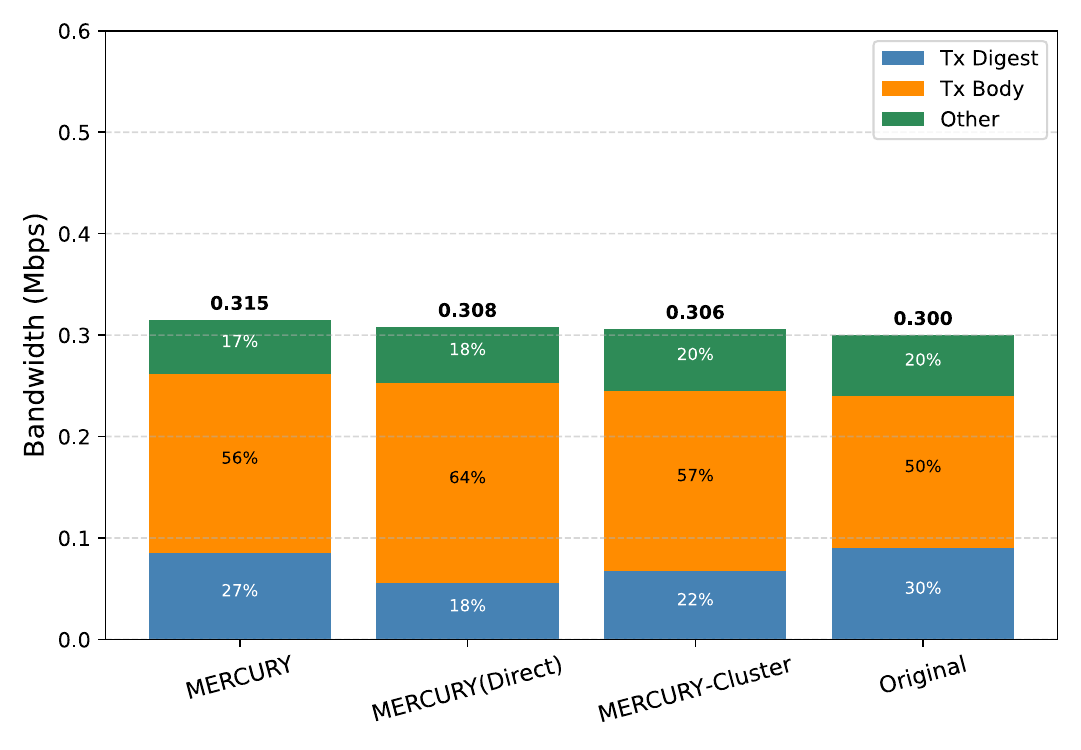}
    \captionof{figure}{Bandwidth breakdown on Conflux.}
    \label{fig:conflux-bw}
  \end{minipage}
\end{figure}

Figure~7 (not shown) confirms that \textsc{BlockSDN-VC} cuts median latency by \textbf{41\%} compared to vanilla Conflux; the 95-th percentile drops from 5.5\,s to 2.3\,s. Sending full bodies in the outburst shaves another 18\% since it avoids the digest/bitmap round-trip.
The latency results are plotted in Fig.~\ref{fig:conflux-cdf}.
``Average Tx $\rightarrow$ packed'' measures the earliest time a transaction appears in a block. \textsc{BlockSDN-VC} improves this metric by \textbf{38\%}, and \textsc{BlockSDN-VC} (Direct) by \textbf{46\%}, confirming that faster propagation directly boosts throughput—sustained TPS rises from 310 (vanilla) to 363 (+17\%).

Figure~8 shows that the additional control and outburst traffic raises total bandwidth by $\approx$5\%—slightly above the 3\% theoretical bound—since each outburst wave includes extra TXD headers ($\approx$40\,bytes) absent in steady-state relays. \textsc{BlockSDN-VC} (Direct) avoids TXDs at the root and thus uses marginally less bandwidth than full \textsc{BlockSDN-VC}, despite sending larger payloads. In all cases, bandwidth stayed well below Conflux's 2\,MB/s anti-spam limit.

In a realistic 12-region deployment, \textsc{BlockSDN-VC} achieves the same gains observed in simulation—$\sim 1.6\times$ faster propagation and $1.4$--$1.5\times$ faster block inclusion—while incurring only modest bandwidth overhead. The direct-send variant offers a further 8--12\% latency drop at nearly identical network cost, making it an attractive drop-in for high-throughput permissioned clusters.

\section{Related Work}

A large body of work proposes alternative topologies for faster dissemination. Kadcast builds a DHT-like overlay whose hop count is logarithmic in the network size, but because it ignores RTT heterogeneity, its latency advantage over random relay is modest. BlockP2P and its variants Urocissa and FRING partition the overlay into geo-clusters and appoint one or two ``routing nodes'' per cluster; these static gateways are single points of failure and become bottlenecks in adversarial settings. CougaR relies on raw RTT probes but offers no defence against forged measurements. Peer-grading strategies—first explored by early Bitcoin clients and later generalized by \textsc{Perigee}’s online-learning UCB rule—let each node drop underperforming peers.

\textsc{BlockSDN-VC} differs in three ways: (i) the VCS is computed by a globally coherent, telemetry-cross-checked SDN controller; (ii) each node retains four uniformly random bridges to preserve random-graph connectivity under cluster sabotage; and (iii) the latency-aware relay list is refreshed in a single control round instead of being learned by trial and error.

\section{Conclusion}

Fast, predictable dissemination is the tightest bottleneck in modern high-throughput blockchains. \textsc{BlockSDN-VC} tackles this problem by lifting the virtual-coordinate computation out of the churn-prone peer layer and into an SDN controller that already monitors the physical network ~\cite{jia2025blocksdn}. The controller’s globally consistent embedding, coupled with a congestion-aware early-outburst relay policy and a small set of random escape edges, trims one full hop from the critical path, reacts to congestion within a single control round, and degrades no worse than random gossip when the controller is offline.

Packet-level simulations over an 8\,000-node Ethereum-sized overlay show that \textsc{BlockSDN-VC} lowers median transaction-propagation latency by up to \textbf{62\%} while adding only \textbf{$\approx$3\%} bandwidth overhead. In a 12-region Conflux deployment, it boosts confirmed throughput by \textbf{17\%} and shortens time-to-pack by \textbf{46\%}, yet withstands coordinate-forgery and black-hole attacks even when nearly half the nodes are malicious.

\section*{Acknowledgements}
This work was funded by the National Key Research and Development Program of China (2022YFB2702300).

%
%
%
%

\end{document}
\bibitem{jia2025blocksdn}Jia, W., Lei, K.: BlockSDN: Towards a High-Performance Blockchain via Software-Defined Cross Networking Optimization. 
In: Proceedings of the 2025 6th International Conference on Computer Engineering 
and Intelligent Control (ICCEIC 2025), 
Guangzhou, China, Oct 17--19, 2025. 


\end{thebibliography}
\end{document}